\documentclass[sort&compress]{elsarticle}
\usepackage{lineno}
\usepackage{amssymb}
\usepackage{amsmath,epsfig}
\usepackage[latin1]{inputenc}
\usepackage{graphicx}
\usepackage[english]{babel}
\usepackage{xspace}
\usepackage{subfigure}
\usepackage{float}
\usepackage{multirow}
\usepackage{hyperref}
\usepackage{placeins}
\usepackage[percent]{overpic}
\usepackage{eurosym}
\usepackage{tablefootnote}
\usepackage{hyperref}
\usepackage{array}
\newcolumntype{L}[1]{>{\raggedright\let\newline\\\arraybackslash\hspace{0pt}}m{#1}}
\newcolumntype{C}[1]{>{\centering\let\newline\\\arraybackslash\hspace{0pt}}m{#1}}
\newcolumntype{R}[1]{>{\raggedleft\let\newline\\\arraybackslash\hspace{0pt}}m{#1}}

\usepackage{color}

\makeatletter
\def\ps@pprintTitle{%
 \let\@oddhead\@empty
 \let\@evenhead\@empty
 \def\@oddfoot{}%
 \let\@evenfoot\@oddfoot}
\makeatother
\journal{Nuclear Instruments and Methods in Physics Research }

\begin{document}

\begin{frontmatter}

\title{Test of Ultra Fast Silicon Detectors for Picosecond Time Measurements with a New Multipurpose Read-Out Board}
\author[add1]{N.~Minafra}
\ead{nicola.minafra@cern.ch}
\author[add1]{H.~Al~Ghoul}
\author[add4]{R. Arcidiacono}
\author[add2]{N. Cartiglia}
\author[add1]{L.~Forthomme}
\author[add2,add3]{R.~Mulargia}
\author[add3]{M. Obertino}
\author[add1]{C.~Royon}

\address[add1]{University of Kansas, Lawrence, USA.}
\address[add2]{INFN, Torino, Italy.}
\address[add3]{Universit\`a di Torino, Torino, Italia and INFN, Torino, Italy}
\address[add4]{Universit\`a del Piemonte Orientale, Italia and INFN, Torino, Italy}

\begin{abstract}
Ultra Fast Silicon Detectors (UFSD) are sensors optimized for timing measurements employing a thin multiplication layer to increase the output signal. A multipurpose read-out board hosting a low-cost, low-power fast amplifier was designed at the University of Kansas and tested at the European Organization for Nuclear Research (CERN) using a 180 GeV pion beam. The amplifier has been designed to read out a wide range of detectors and it was optimized in this test for the UFSD output signal. In this paper we report the results of the experimental tests using 50 $\rm{\mu m}$ thick UFSD with a sensitive area of 1.4 $\rm{mm^2}$. A timing precision below 30 ps was achieved. 
\end{abstract}


\end{frontmatter}

%\linenumbers

\section{Introduction}
\label{introduction}
The increasing demand for better timing and spatial resolution has been the key purpose for developing new type of sensors.
Low Gain Avalanche Detector (LGAD) technology is one of the most promising advancements in silicon detector technology as it allows a sensible increase of the detector output signal, while keeping all of the known advantages of a silicon substrate, such as low cost and large scale production capabilities.
A sensor with an enhanced output signal can also be used for tracking since a large signal simplifies the design of the front-end electronics; however, its most innovative application is for precise timing.
When a particle detector is crossed by a minimum ionizing particle (MIP), free charge carriers are produced inside the sensitive region. An output signal is then produced when these charge carriers start drifting towards the electrodes where they will be collected.
The duration of the signal depends on the length of the collection process; therefore a thin sensor produces a faster signal. On the other hand, a thin sensor has a small sensitive volume and, therefore, produces a smaller signal.
With the LGAD technology it is possible to create a thin sensor that produces a signal 10 to 20 times larger than the equivalent traditional sensor.
Thin LGAD sensors optimized for precise time measurements, also known as UFSD, have been developed by CNM \cite{cnm_ufsd}.
%A new UFSD design with 50 $\rm{\mu m}$ thick sensors has been developed by CNM Barcelona and optimized for precise time measurements \cite{cartiglia2015design}.

A multipurpose read-out board has been developed to test different type of sensors and it has been optimized in this test for the UFSD characteristics. 
%The board is part of the effort of the University of Kansas to build a test laboratory to test solid state detectors, especially focusing on time performance. 
This board can be used to perform a full characterization of the sensor, i.e. current/voltage (\emph{IV}) characteristics, capacitance measurement, test with pulsed lasers, radioactive sources and particle beams.

In section \ref{simulation} the multipurpose read-out board will be described, with particular focus on the optimization of the parameters of the amplifier to 50 $\rm{\mu m}$ thick UFSD.
The following section describes how the board was used to characterize the sensor in the North Area at CERN.
Finally, in section \ref{results}, the results are presented and discussed.

\section{The multipurpose read-out board}
\label{simulation}
In this section, the multipurpose read-out board will be described: first the layout of the board is discussed, then the optimization process for a particular detector, using numerical simulations, is illustrated.

The read-out board has been designed to host solid state sensors of different types such as UFSD, traditional silicon detectors, Silicon PhotoMultipliers (SiPMs) as well as diamond detectors. 
The board
%, that will be submitted for a patent, 
includes an amplifier that can be configured to have a very large bandwidth, for instance when reading SiPMs or high gain Avalanche PhotoDiodes (APDs), or to optimize the signal-to-noise ratio (SNR) when reading diamond detectors and LGADs.

\begin{figure}[!h]
\centerline{%
\includegraphics[width=0.7\textwidth]{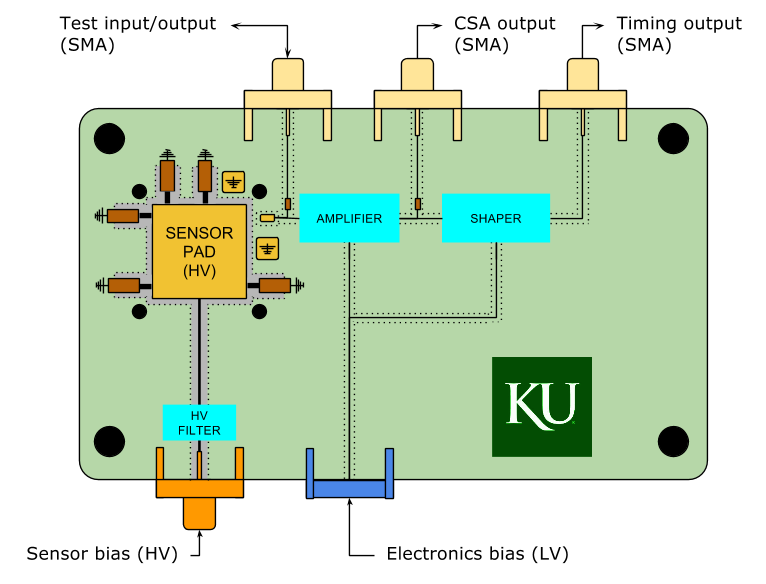}}
\caption{Layout of the front-end board. Different connectors are used for the sensor, electronics bias, signal inputs and outputs. }
\label{Fig:Drawing}
\end{figure}

The board can distribute a bias potential up to 1 kV through a 20x20 $\rm{mm}^2$ pad on which the sensor is glued. Grounded pads are also available to connect a sensor's guard ring.
Finally, the amplifier is DC-coupled to the sensor.
The test board allows reading out the signal at various stages, as shown in Fig. \ref{Fig:Drawing}:
\begin{itemize}
\item directly from the sensor;
\item after a low noise charge sensitive amplifier (CSA);
\item after a shaping stage optimized for time measurements.
\end{itemize}
The first connector can also be used to send a calibration signal to the amplifier.

One of the main contributions to the time precision of a detector is the jitter, which can be approximated to\footnote{An extended discussion of the different contributions that affect a time measurements is done in chapter 6 of~\cite{Minafra_PhD}.}:

\begin{linenomath}
\begin{equation}
\sigma_t \approx \sigma_{\mathrm noise}\cdot\frac{\tau}{A} = \frac{\tau}{SNR} = 1.25 \frac{\tau_{\mathrm{0.1-0.9}}}{SNR},
\label{eq:cfd}
\end{equation}
\end{linenomath}
where $\tau$ is the rise time of the signal, $\tau_{\mathrm{0.1-0.9}}$ is the 10\% to 90\% rise time; A is the amplitude and $\sigma_{\mathrm noise}$ is the RMS of the noise.
In this approximation, a large SNR with a slow signal or a small SNR with a fast signal are equivalent.
%However, in real scenarios, the best compromise depends on the sensor and on the application.

The electronics can be optimized for different sensors using numerical simulations, starting from a model of the sensor. In particular, it is possible to increase or decrease the gain of the last stage of amplification (with a negligible effect on the SNR) or to change the time constant of the shaper: a larger time constant is needed to produce a slower signal with improved SNR.

In this work, the amplifier parameters were optimized for 50 $\rm{\mu m}$ UFSDs, for MIP detection and for a signal read-out using a fast sampling device, like an oscilloscope or the SAMPIC chip \cite{breton2016measurements}. Thanks to the large signal produced by the sensor, the time constant of the amplifier was reduced below 1 ns, enough to have a good SNR. Then, the final stage was tuned to produce a signal amplitude (for the MIP) around 100 mV, to cope with the dynamic range of the SAMPIC chip.

\begin{figure}[!h]
\centerline{\includegraphics[width=0.8\textwidth]{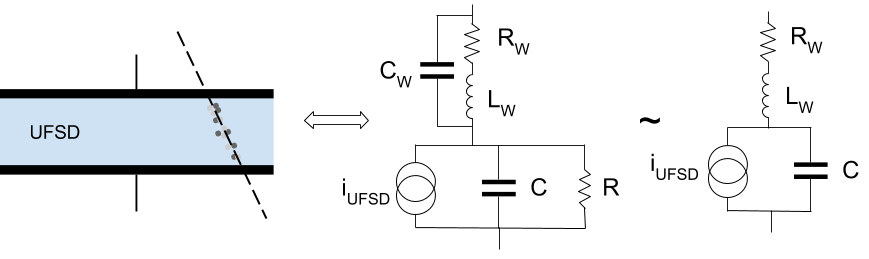}}
\caption{Sketch of the sensor equivalent circuit. The sensor can be simulated as a current generator in parallel with a capacitor and connected to the amplifier through an inductor and a resistor.}
\label{Fig:UFSD_model}
\end{figure}

The equivalent model of the sensor used to simulate the electronics is a current generator, in parallel with a capacitor and a resistor. This resistor can be neglected if the dark current is low, as shown in Fig. \ref{Fig:UFSD_model}. It is also possible to include the parasitic effects of the input connection, i.e. bonding wire, bonding pad, and path to the amplifier among others.
The dominant effects are due to the capacitance of the sensor and to the inductance and resistance of the bonding wire.

The current produced by the detector was simulated using \texttt{Weightfield2} \cite{weightfield2} and an analytic replica was used as input to the amplifier simulation.
This approach works for all sensors simulated by the software, i.e. silicon and diamond detectors and UFSD of any given thickness and geometry.  

% \begin{figure}[!h]
% \centerline{%
% \includegraphics[width=0.52\textwidth]{img/signal_weightfield}
% \includegraphics[width=0.47\textwidth]{img/signal_spice}}
% \caption{The signal generated by the sensor has been simulated using \texttt{Weightfield2} \citep{weightfield2} and reproduced in an analytic form in the amplifier simulation.}
% \label{Fig:signal_weightfield}
% \end{figure}

\begin{figure}[!h]
\centerline{\includegraphics[width=\textwidth]{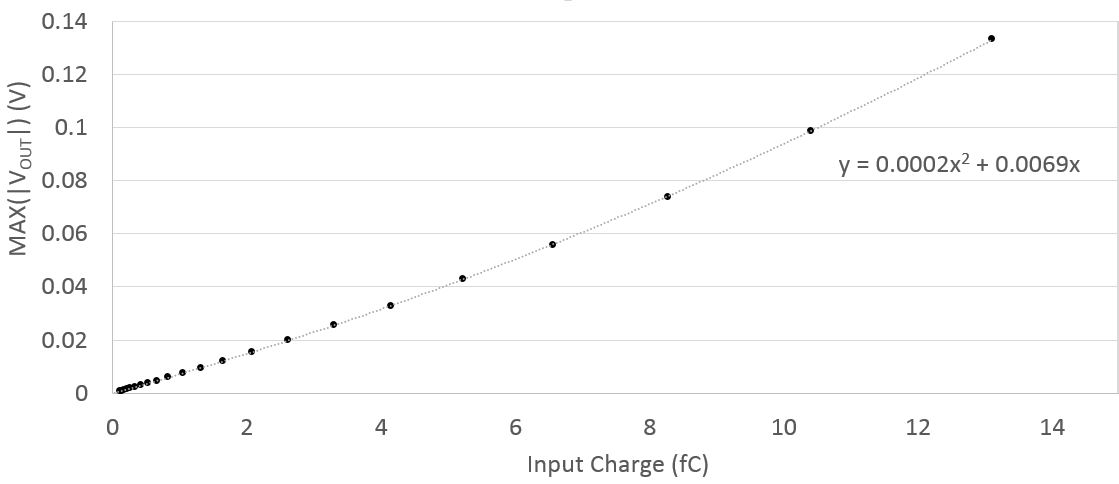}}
\caption{The simulated output of the amplifier connected to a 50 $\rm{\mu m}$ UFSD can be assumed linear for small input charges ($<4~\rm{fC}$), equal to $\approx~7~\rm{mV/fC}$ while for higher values a parabola is needed.}
\label{Fig:Amplifier_gain}
\end{figure}

Fig. \ref{Fig:Amplifier_gain} shows the expected output amplitude for a 50 $\rm{\mu m}$ UFSD.
The gain for low input charge is $\approx~7~\rm{mV/fC}$, while the expected noise RMS is $\approx~0.23~\rm{fC} \approx 1400 e^-$.

\begin{figure}[!h]
\centerline{%
\includegraphics[width=\textwidth]{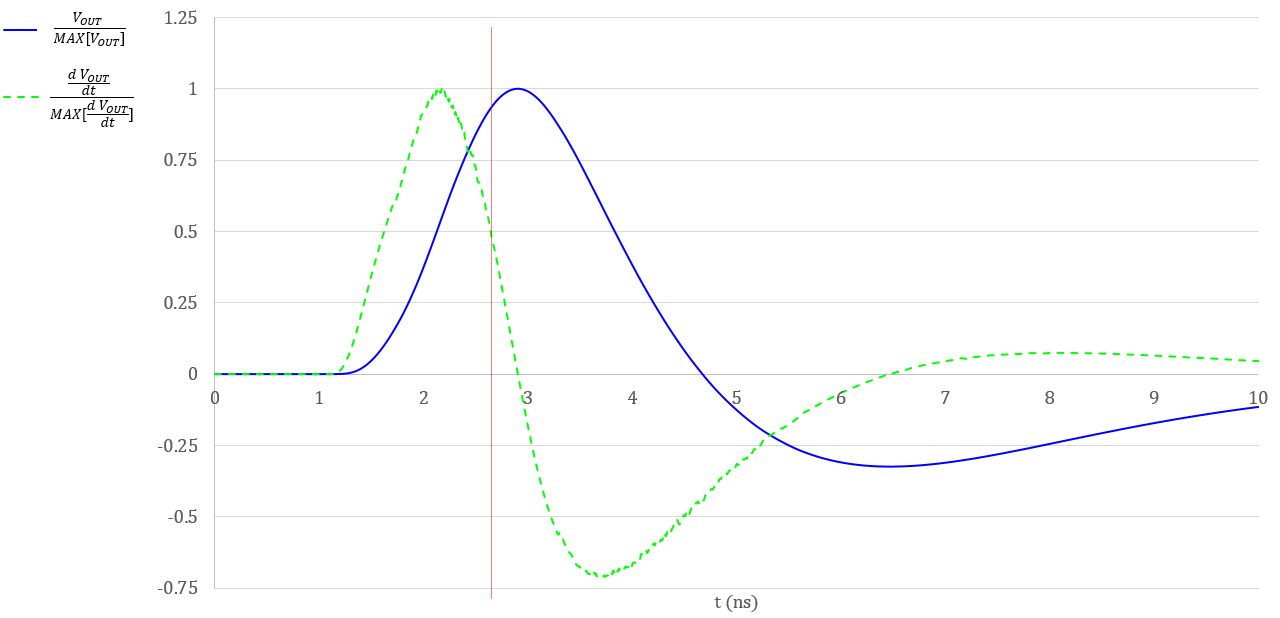}}
\caption{The signal slope is maximum at 50\% of the maximum amplitude.}
\label{Fig:CFD_Simulated}
\end{figure}

The simulation of the electronics also gives suggestions on the parameters to be used to minimize the jitter. For example, using a constant fraction discriminator the best performance is obtained using a threshold at 50\% of the amplitude, as the derivative of the signal is maximum at that point as shown in Fig. \ref{Fig:CFD_Simulated}.

% After the optimization of the electronics, the board has been manufactured; and the sensor has been glued and bonded to the board.

% \FloatBarrier
\section{Test on particle beam}
\label{setup}
The board, i.e. Device Under Test (DUT), was installed on a test bench at CERN in the H8 experimental area on an SPS secondary beam.
In this section, the test setup will be described.

The signal was read using an oscilloscope (\emph{Agilent} DSO9254A) acquiring at 20 GS/s. The time precision was measured using as a reference a 10 mm long quartz bar of 9 $\rm{mm^2}$ glued on a \emph{SensL} Silicon Photomultiplier (SiPM) of the same surface, installed on an evaluation board of the same manufacturer \cite{sensl_evboard}. 

An additional detector, a quartz bar installed on a MCP-PMT \emph{Planacon} XP85012, was used to cross check the measurement of the time precision.
By measuring the time difference between each combination of two detectors, it is possible to uniquely compute their time precision, using the setup shown in Fig. \ref{Fig:ExpSetup}:

\begin{figure}[!h]
\centerline{%
\includegraphics[width=0.55\textwidth]{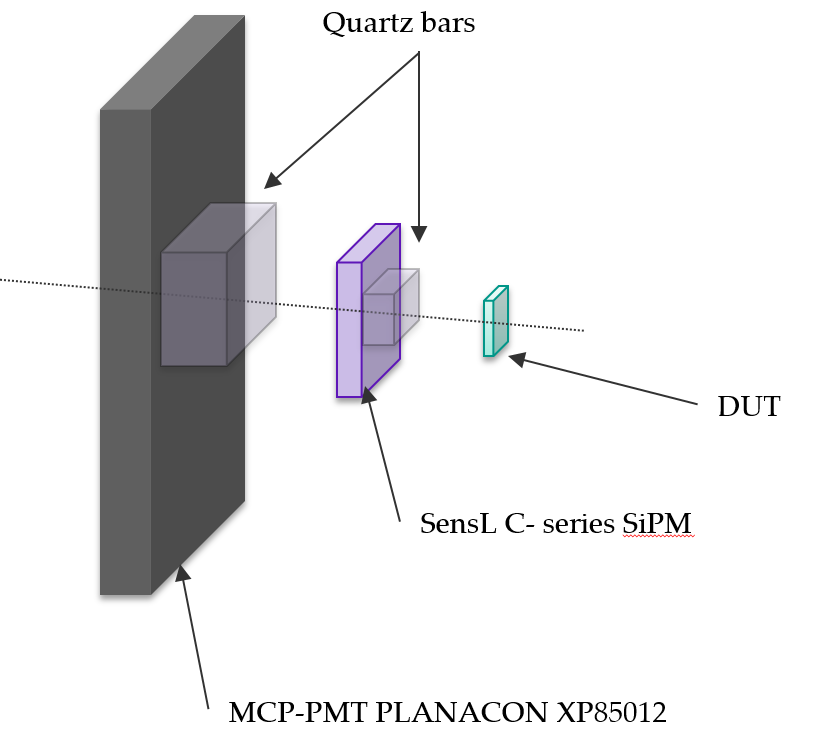}}
\caption{Schematic of the experimental setup. A SiPM was used as time reference and a MCP-PMT was used to measure the time precision of the SiPM. }
\label{Fig:ExpSetup}
\end{figure}

\begin{equation}
\begin{cases}
\sigma_{\mathrm DUT}^2 + \sigma_{\mathrm SiPM}^2 = \sigma_{\mathrm SiPM-DUT}^2	\\
\sigma_{\mathrm SiPM}^2 + \sigma_{\mathrm MCP}^2 = \sigma_{\mathrm MCP-SiPM}^2	\\
\sigma_{\mathrm MCP}^2 + \sigma_{\mathrm DUT}^2 = \sigma_{\mathrm MCP-DUT}^2,	\\
\end{cases}
\end{equation}
where $\sigma_{\mathrm DUT}$ is the time precision of the DUT, $\sigma_{\mathrm SiPM}$ is the time precision of the SiPM and $\sigma_{\mathrm MCP}$ is the time precision of the MCP-PMT; while $\sigma_{\mathrm i-j}$ are the standard deviation of the measured time difference between device $i$ and $j$. 
Hence the individual time precisions are given by
\begin{equation}
\begin{cases}
\sigma_{\mathrm DUT}^2 = ( \sigma_{\mathrm SiPM-DUT}^2 - \sigma_{\mathrm MCP-SiPM}^2 +\sigma_{\mathrm MCP-DUT}^2 ) / 2 	\\
\sigma_{\mathrm SiPM}^2 = ( \sigma_{\mathrm MCP-SiPM}^2 -\sigma_{\mathrm MCP-DUT}^2 + \sigma_{\mathrm SiPM-DUT}^2 ) / 2	\\
\sigma_{\mathrm MCP}^2  = ( \sigma_{\mathrm MCP-DUT}^2 - \sigma_{\mathrm SiPM-DUT}^2 + \sigma_{\mathrm MCP-SiPM}^2 ) / 2 	\\
\end{cases}
\end{equation}

Using these equations, the time precision of the SiPM was estimated to be $\approx 18~\rm{ps}$ and the precision of the MCP-PMT $\approx 22~\rm{ps}$. The error on these measurements was estimated to be of the order of 5\%. 

\begin{figure}[!h]
\centerline{%
\includegraphics[width=0.7\textwidth]{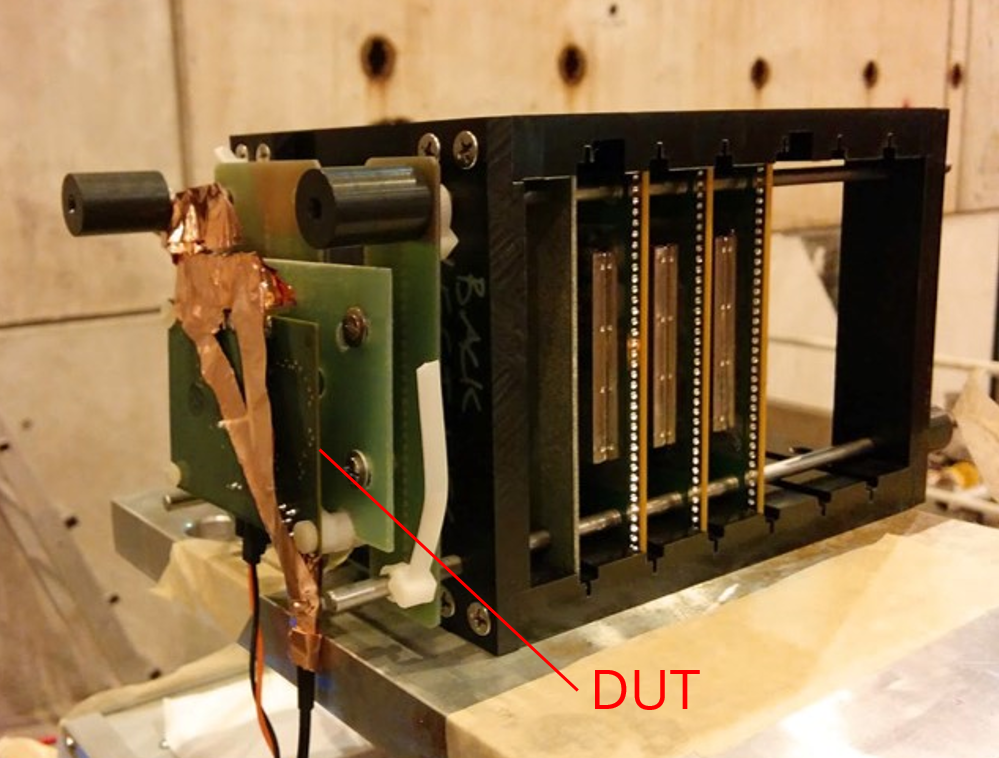}}
\caption{Picture of the experimental setup. The DUT is installed on a support structure to ease the alignment. }
\label{Fig:ExpSetupPicture}
\end{figure}

The alignment of the detectors was insured by a mechanical structure, shown in Fig. \ref{Fig:ExpSetupPicture}.

\FloatBarrier
\section{Beam test results}
\label{results}
Data acquired during the beam test have been analyzed and the results will be presented and discussed in this section.

\begin{figure}[!h]
\centerline{%
\includegraphics[width=.95\textwidth]{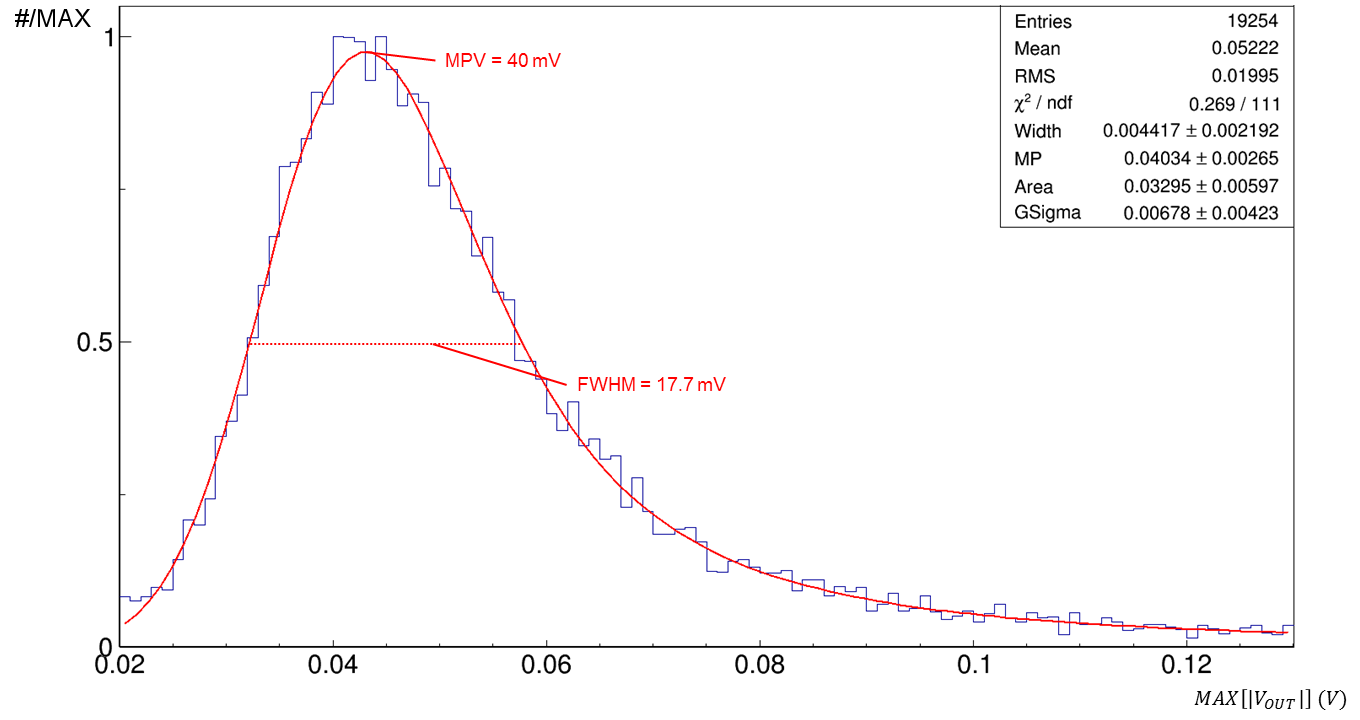}}
\caption{Amplitude of the amplified signal produced by the UFSD biased at 180~V. The distribution obtained convolving a Landau with a Gaussian that best fits the data and its parameters are also shown. }
\label{Fig:Landau_180V}
\end{figure}

One may investigate the performance of the sensor at different gain values. The latter can be controlled by increasing the bias voltage \cite{sadrozinski2014sensors}.
The amplitude of the output signal depends, for a fixed gain, on the charge deposited by the passage of the particle; therefore, the distribution of the signal amplitude should follow a Landau distribution. 
However, if noise is added to the signal, the amplitude will be distributed as a Landau convoluted with a Gaussian. The parameters of the Landau distribution are the most probable value (MPV) and the width, that is $1/4$ of the Full Width Half Maximum (FWHM); the $\sigma$ of the Gaussian ($GSigma$) is an estimation of noise and fluctuations.
This convolution fits the data very well as it can be seen in Fig. \ref{Fig:Landau_180V}, showing the amplitude distribution at a bias voltage of 180~V. 

\begin{figure}[!h]
\centerline{%
\includegraphics[width=.9\textwidth]{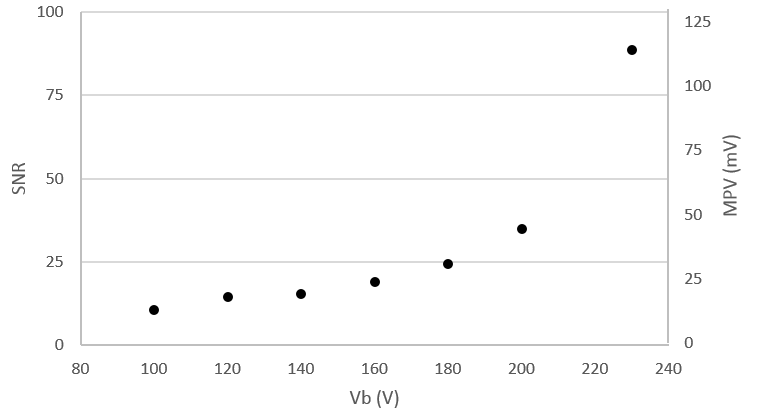}}
\caption{Dependence of the measured signal amplitude (Most Probable value) and of the SNR increasing the biasing voltage Vb, computed considering only the noise contribution of the amplifier that is constant $\sigma_v \approx 1.5~\rm{mV}$.}
\label{Fig:Ufsd_SNR_vs_vb}
\end{figure}

The MPV depends on the gain, hence the bias voltage, while the amplifier noise is independent from the latter. The SNR has the same dependence of the MPV, as shown in Fig. \ref{Fig:Ufsd_SNR_vs_vb}.

The term $GSigma$ has two main contributions, from the sensor, i.e. shot noise, and from the amplifier: $GSigma^2 \approx \sigma_{\mathrm sensor}^2 + \sigma_{\mathrm amplifier}^2$. The contribution of the amplifier can be measured in absence of particles (fluctuation of the pedestal). It was measured to be constant even among different acquisitions taken with different environmental factors (temperature, radiofrequency pick-ups, etc).
% On the other hand, the contribution of the sensor increases with the gain, as shown in Fig. \ref{Fig:GSigma_vs_vb}.

% \begin{figure}[!h]
% \centerline{%
% \includegraphics[width=.9\textwidth]{img/GSigma_vs_vb}}
% \caption{The parameter $GSigma$ is an indication of the noise produced by the sensor that, as expected, becomes larger increasing the bias voltage.}
% \label{Fig:GSigma_vs_vb}
% \end{figure}

\begin{figure}[!h]
\centerline{%
\includegraphics[width=.9\textwidth]{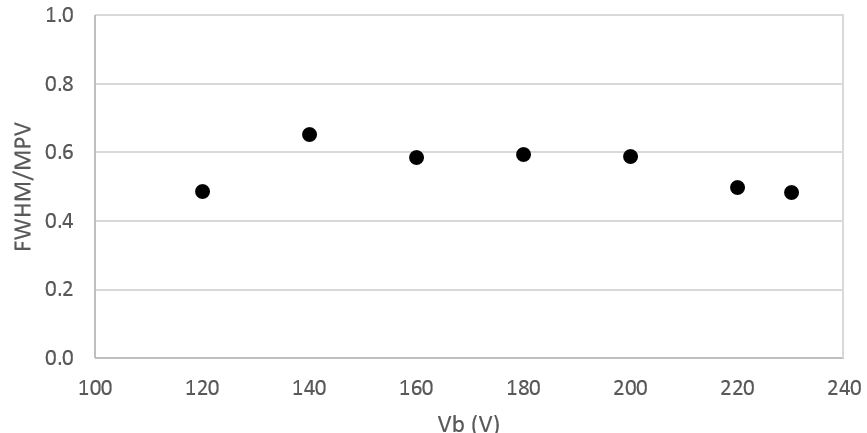}}
\caption{The stochastic behaviour of the signal amplitude can be described by the Landau distribution. The ratio between the Full Width at Half of the Maximum and the Most Probable Value of this distribution is expected to be close to 0.6 for a $50~\rm{\mu m}$ thick silicon sensor \cite{FWHM_ForSilicon}.}
\label{Fig:Ufsd_fwhm_vs_vb}
\end{figure}

The behaviour of the ratio between the MPV and the Full Width at Half of the Maximum (FWHM) of the amplitude distribution\footnote{These values are computed directly from the histogram of the amplitude, without using a Landau distribution, i.e. without deconvolving the noise.} was investigated. As shown in Fig. \ref{Fig:Ufsd_fwhm_vs_vb}, this ratio mainly depends on the thickness of the sensor and the value is expected to be around 0.6 for a $50~\rm{\mu m}$ thick silicon sensor \cite{FWHM_ForSilicon}.
% The measured values are below the expected ratio as the FWHM of the Landau distribution that best fits the data is an underestimation of the real value, as noticeable in Fig. \ref{Fig:Landau_180V}.

\begin{figure}[!h]
\centerline{%
\includegraphics[width=.9\textwidth]{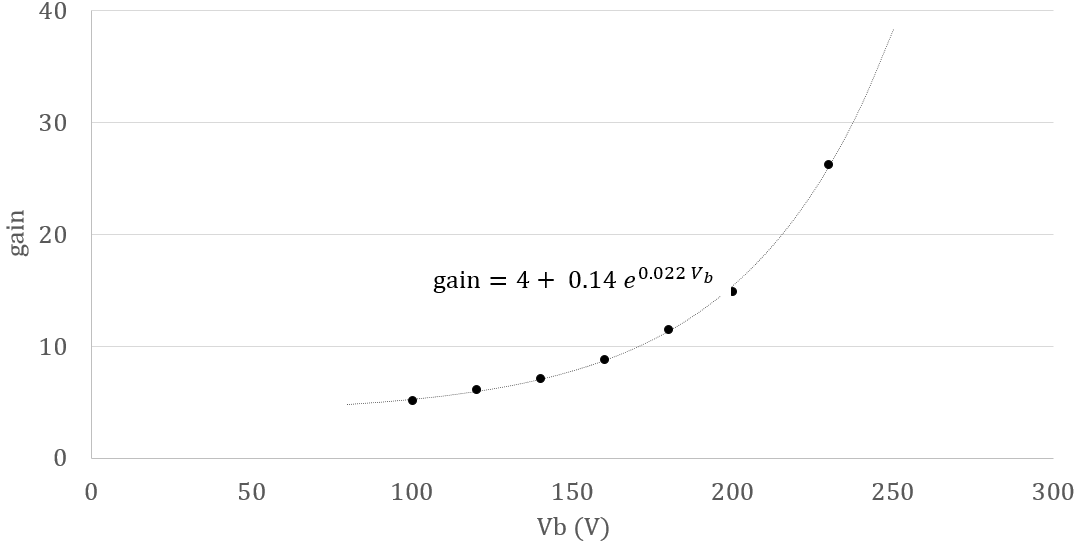}}
\caption{Dependence of the gain of the UFSD on the applied biasing voltage. The black dots represent the measured values and the equation of the exponential function that best fits the data is also shown.}
\label{Fig:gain_fit}
\end{figure}

\begin{figure}[!h]
\centerline{%
\includegraphics[width=.9\textwidth]{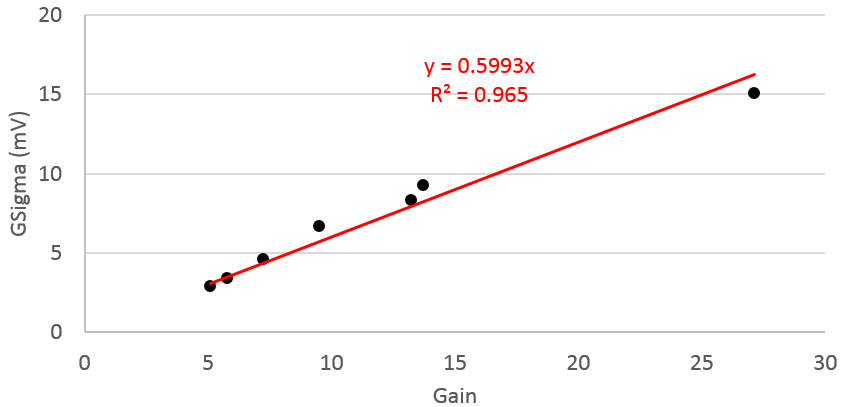}}
\caption{The parameter $GSigma$ is an indication of the noise produced by the sensor and increases linearly with gain. The red solid line is a linear interpolation obtained requiring $GSigma=0$ with no gain.}
\label{Fig:GSigma_vs_gain}
\end{figure}

The gain of the sensor is investigated measuring the amplitude of the signal for different bias voltages, normalized to the simulated amplitude for an equivalent sensor without the gain layer (gain = 1). The results are shown in Fig. \ref{Fig:gain_fit} and are in good agreement with the results found in other works, for example \cite{cartiglia2016beam}.
Fig. \ref{Fig:GSigma_vs_gain} shows that the noise $GSigma$ increases linearly with the gain.

\begin{figure}[!h]
\centerline{%
\includegraphics[width=.9\textwidth]{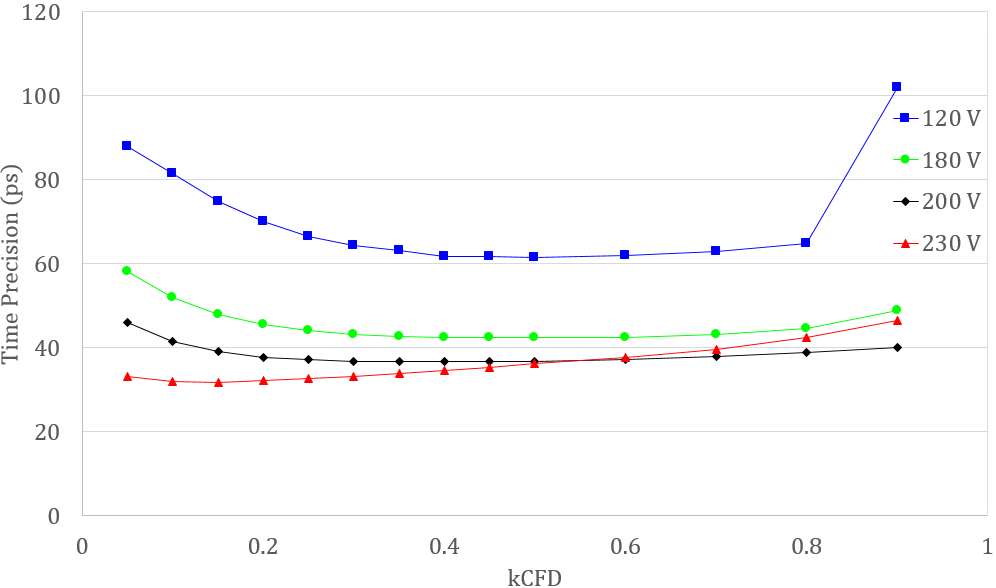}}
\caption{Dependence of the measured time precision on the fraction used for the CFD. The best CFD ratio is 0.5, as expected from simulations; however, when the biasing voltage is high, i.e. high gain, the behaviour is different.}
\label{Fig:kCFD_dep}
\end{figure}

Finally, using a SiPM as time reference (see section \ref{setup}), the precision of the measured time of arrival was computed for different settings, using an off-line Constant Fraction Discriminator (CFD).
The dependency on the fraction used in the CFD ($k_{\mathrm{CFD}}$) is shown in Fig. \ref{Fig:kCFD_dep} for various bias voltages. As discussed in section \ref{simulation}, the best ratio to minimize the jitter is around 50\%, while for higher gain, where the jitter is negligible, a lower value is preferable.

\begin{figure}[!h]
\centerline{%
\includegraphics[width=.9\textwidth]{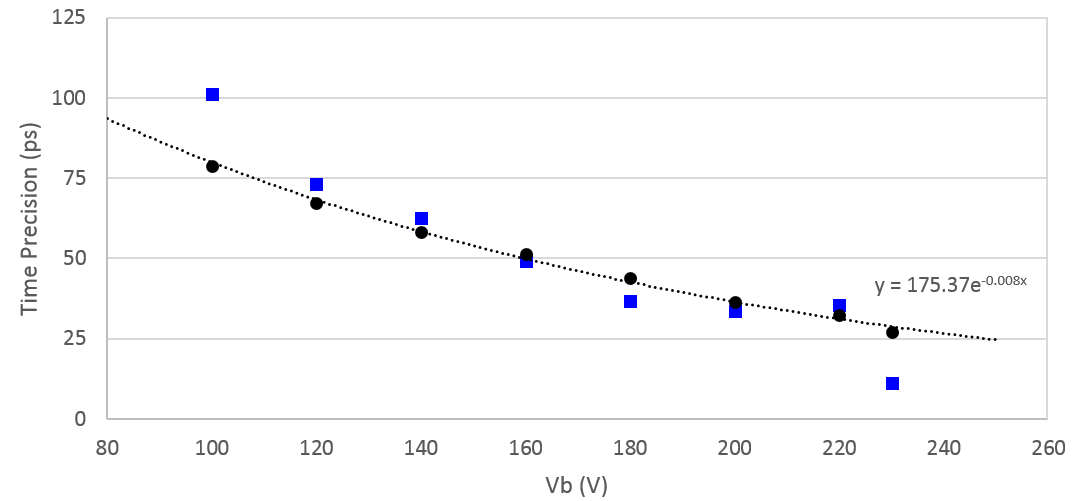}}
\caption{Dependence of the measured time precision on the biasing voltage. The black circles are the measured values using the best value of $k_\mathrm{{CFD}}$, while the blue squares are the time precision computed using eq. \ref{eq:cfd} considering only the noise of the amplifier.}
\label{Fig:Ufsd_tp_vs_vb}
\end{figure}

As shown in Fig. \ref{Fig:Ufsd_tp_vs_vb}, the time precision is improved increasing the bias voltage. A deviation from the precision expected using eq. \ref{eq:cfd} is noticeable and the deviation can be attributed to the non linear rising edge of the signal and to the Landau noise \cite{cartiglia2016tracking}.

\begin{figure}[!h]
\centerline{%
\includegraphics[width=.95\textwidth]{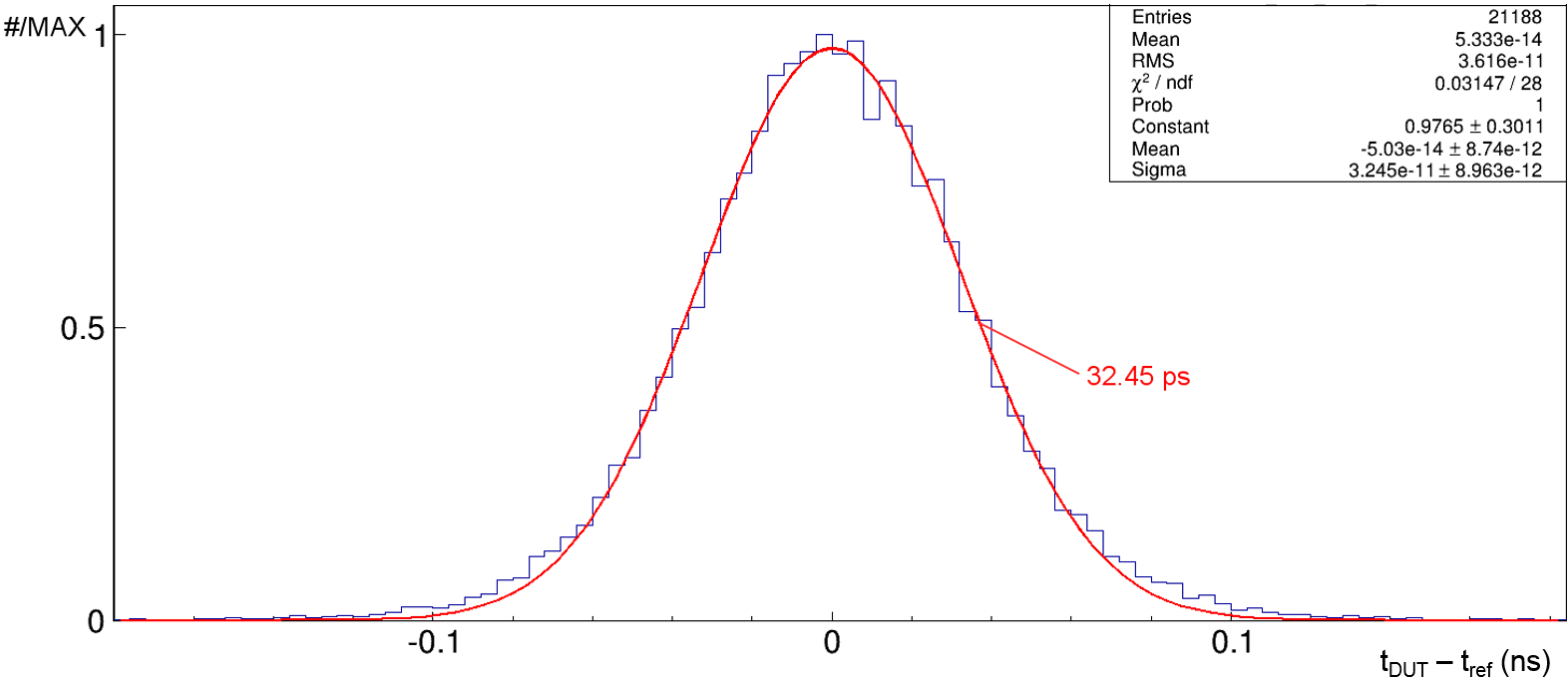}}
\caption{Time difference between the UFSD under test, biased at 230~V, and the time reference, time precision of the UFSD is $\approx \sqrt{\sigma^2 - \sigma^2_{\mathrm ref}} \approx 27~\rm{ps}$}
\label{Fig:Sigma_230V_BestCorrected}
\end{figure}

Fig. \ref{Fig:Sigma_230V_BestCorrected} shows the difference between the time of arrival measured using the UFSD under test and the time reference.

The data presented in this section indicate that the multipurpose board is an excellent tool for the characterization of solid state sensors and, in combination with the latest generation of UFSD, can reach a time precision below 30 ps for MIPs.

\section{Conclusion}
\label{conclusion}
The design, realization, and test of a general purpose front-end board for solid state detectors was discussed in this work. The board was specifically designed for precise time measurements using a sampling device, such as a digital oscilloscope or a SAMPIC chip. It was tested using a 50 $\rm{\mu m}$ thick UFSD and a time precision down to 27 ps was measured with a pion beam in the North Area at CERN.
A more compact and low power version of the amplifier is under development and is specifically designed for applications where compact electronics are needed but the development of a dedicated integrated circuit is not feasible for time nor cost constraints.

\section{Acknowledgements}
The authors would like to thank the TOTEM Collaboration for generously  giving up part of their beam time in the North Area at CERN to allow performing the tests discussed in this paper and also the RD50 Collaboration for the sensors production and UCSC for the time reference.

The work was supported by the United States Department of Energy, grant DE-FG02-04ER41286. Part of this work has been financed by the Spanish Ministry of Economy and Competitiveness through the Particle Physics National Program (FPA2015-69260-C3-3-R and FPA2014-55295-C3-2-R), by the European Union's Horizon 2020 Research and Innovation funding program, under Grant Agreement no. 654168 (AIDA-2020) and Grant Agreement no. 669529 (ERC UFSD669529), and by the Italian Ministero degli Affari Esteri and INFN Gruppo V.

\section*{References}
\bibliography{biblio}

\end{document}